# Halley's Wind: Reappraising a Centuries Old Theory for the Trade Winds

Kristopher B. Karnauskas [1, 2, *]

[1] Department of Atmospheric and Oceanic Sciences, University of Colorado Boulder
[2] Cooperative Institute for Research in Environmental Sciences, University of Colorado Boulder
* Corresponding author (kristopher.karnauskas@colorado.edu)

## Abstract

The astronomical work of Edmond Halley (1656–1742) during the latter half of the Copernican Revolution contributed substantially to the scientific proof and widespread acceptance of heliocentrism. Perhaps the most well–known example is his prediction of the return of the comet named in his honor. Halley is also known for offering an incorrect account of Earth's atmospheric general circulation in 1686, specifically the cause of the easterly trade winds in the tropics. In light of updated data and models, the notion of the 'wrongness' of Halley's proposed explanation is reconsidered when cast within the milieu of the prevailing cosmological model. The key ingredient to Halley's mechanism, underappreciated at the time by Halley and his contemporaries, is the thermal inertia of the climate system. It has been suggested by some authors that Halley's ideas on atmospheric circulation were the first to "go beyond the Ptolemaic view of climate." Ironically, Halley's explanation was essentially correct, but only in a Ptolemaic (geocentric) universe.

## Main text

Famous for the comet named after him, Edmond Halley made groundbreaking discoveries in astronomy and helped seal the Copernican Revolution that the Earth revolved around the Sun, not vice versa. Less well known is that Halley also put forth pioneering ideas in the early 1700s about how the Earth's atmosphere works. Those ideas have long been declared dead wrong, but a modern–day look at the three–century–old theory indicates it had some scientific merit.

Nicolaus Copernicus (1473–1543) instigated the revolution in our understanding of Earth and its place in the Universe with his publication of *De Revolutionibus* in 1543 (Fig. 1). By proposing that Earth revolves around the Sun (heliocentrism), Copernicus was upending the Earth–centered (geocentric) model of the cosmos put forth by Ptolemy (90–168) in the $2^{nd}$ Century.

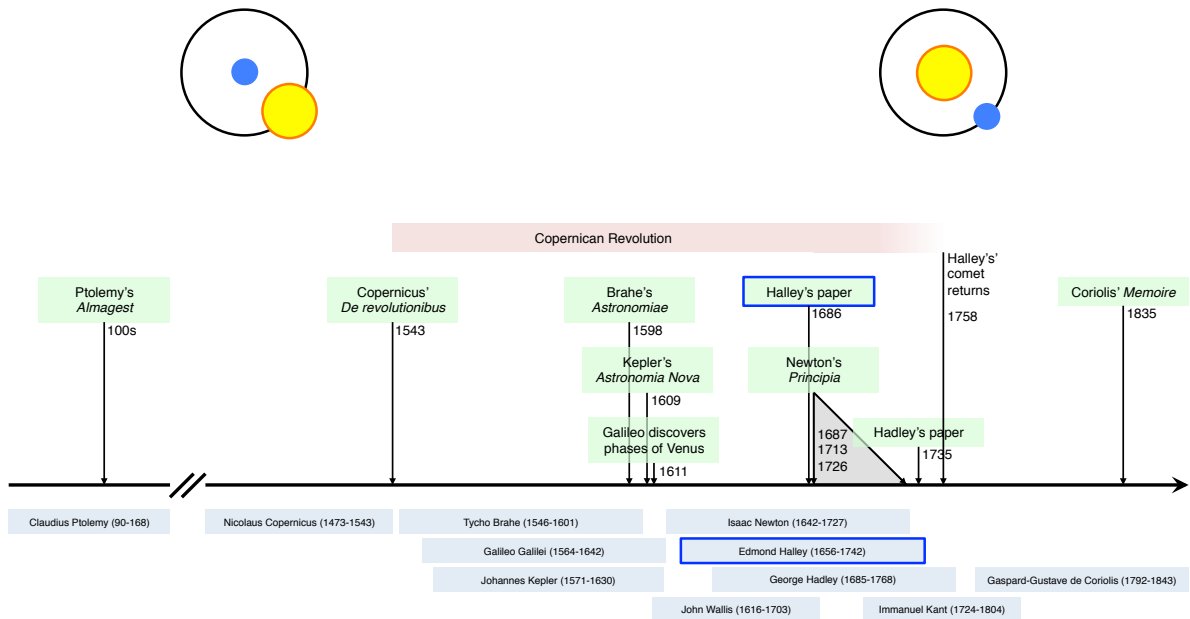

**Figure 1.** Schematic timeline of the key people and dates concerning the development of explanations for the trade winds and cosmological models.

Over the course of the next 200 years, several key figures contributed to the eventual acceptance of the heliocentric model. Among the notable: Johannes Kepler (1571–1630) calculated that the motion of the stars and planets are best explained by elliptical orbits, Galileo Galilei (1564–1642) developed a telescope capable of observing that Venus has phases like our Moon, and Halley (1656–1742) predicted the return of a comet using Isaac Newton's (1642–1727) elegant laws of motion and gravitation – none of which would be possible in a geocentric model. These men were among the most proficient and outspoken supporters of Copernicus' new worldview, and not all of the most brilliant scientists and philosophers of the day were on board (Danielson and Graney 2014). If Newton's *Principia* put the lid on Ptolemy's coffin, the return of Halley's comet in 1758 was the final nail.

Between existential discoveries in astronomy and pioneering developments in technology, both Galileo and Halley also took aim at one of the most important terrestrial puzzles of the day. The winds in the tropics had long been observed to blow steadily from east to the west, commonly referred to as the easterly trade winds. Despite the tremendous societal relevance of this phenomenon – maritime navigation – we were without explanation.

This is perhaps no surprise, since we are now aware that this phenomenon depends critically on the fact that Earth spins on its axis – a contentious idea at the time. Ahead of the curve, Galileo made the assumption that the Earth indeed spins, but it spins eastward so fast that the layer of air hovering over the surface is unable to keep up, thus the wind appears to blow westward relative to an observer standing on the rapidly–spinning planet.

A lifetime later, Halley published a different explanation in 1686 in the world's first scientific journal, *Philosophical Transactions of the Royal Society*. The Sun's rays cast warmth upon half of the Earth (the day side), he said, and that warmth travels westward across the Earth's surface to return again in 24 hours. Presuming that warm air is more "rarefied" (buoyant), there must be rising air over the sunlit patch of Earth. Likewise, over the dark and cold side, the air must be more "ponderous" (heavy) and sink downward. With such disequilibrium, horizontal air currents must be rushing between the dark side and the daytime side. With the patch of solar heating ever–circling westward around the planet, we have the easterly trade winds.

Halley's 1686 paper was very significant for three reasons. First, this was one of the first times the connection was made between differences in solar heating across the Earth's surface (gradients) and atmospheric circulation. Second, this was the first time the atmosphere was said to have a *circulation* – continuous and interconnected circuits of air driven by heating gradients. Third, this was one of the most famous misconceptions in the history of atmospheric science.

Halley's theory gained some traction at the time because the physics behind the idea seem almost intuitive, and perhaps because the model did not demand *a priori* acceptance of a spinning Earth. The sunlight cast upon Earth's surface would move westward in multiple cosmological models; for example, the Sun revolving around a perfectly stationary Earth once daily (Ptolemy's model), or the Earth revolving around the Sun very slowly (once per year) but spinning about its own axis once daily (Copernicus' model).

George Hadley (1685–1768) published the correct (or nearly so) explanation in 1735 for the easterly trade winds – which *does* demand complete adherence to heliocentrism including the spinning Earth – but that went largely unnoticed for several decades. Any mention of Halley's explanation today, from Wikipedia to advanced college textbooks (Gill 1982), inevitably characterizes Halley's explanation for the easterly trades as incorrect, while the correct explanation "fell to another Englishman, George Hadley, who nominated the Earth's rotation as being responsible for altering the circulation" (Stevens 1999).

But is there some merit to Halley's theory? Is there a universe or cosmological model in which Halley's proposition would be correct? What sort of tweaks to the Sun–Earth dynamic as we now understand it would be necessary for Halley's mechanism to be the *only* driver of east–west winds on Earth and for them to be predominantly easterly?

A logical starting point is to remove the mechanism that actually creates the easterly winds – that is, the fact that Earth spins on its axis. Since Halley's mechanism requires the sunlight to propagate westward across the surface of the Earth, let the Sun revolve around the Earth on a daily basis. Even now, however, we find ourselves facing the same objection as the one raised by John Wallis (1616–1703) and later by Immanuel Kant (1724–1804) – that in addition to air rushing in toward the warm, sunlit region from the east, air would also be rushing in from the west (*i.e.*, on the morning side). When Wallis expressed this concern to Halley, Halley himself began to doubt his own theory (Persson 2006).

The key ingredient to Halley's mechanism, underappreciated at the time by Halley and his contemporaries, was the thermal inertia of the climate system. Most people can recount experiencing the rapid warm–up upon sunrise compared to the way the afternoon's warmth can seem to linger well after sunset. This is especially evident over the ocean, where the upper layer of water warms slowly in response to solar radiation and releases the accumulated heat even more slowly into the atmosphere throughout the night. The result is an asymmetric daily cycle of temperature.

Seven years of continuous hourly temperature measurements at 155°W, 0°N, the very center of the largest ocean on Earth, the Pacific, show this playing out; air temperature ascends from nighttime minimum to daytime maximum in just a few hours, but takes about four times as long to cool back down to the minimum (Fig. 2). Clearly, surface temperature is not a simple linear function of incident solar energy on short time scales, as Halley and his contemporaries likely imagined. Assuming for simplicity that the Earth is entirely dominated by ocean, the daily temperature cycle can easily be translated into a profile of temperature along a line of latitude around the globe (with every hour representing 15° longitude or about 1,650 km). Schematically, the temperature distribution at a given moment in time might look like Fig. 3. Lagging just behind the peak solar energy are the warmest temperatures, followed by a greatly elongated region across which temperature decreases.

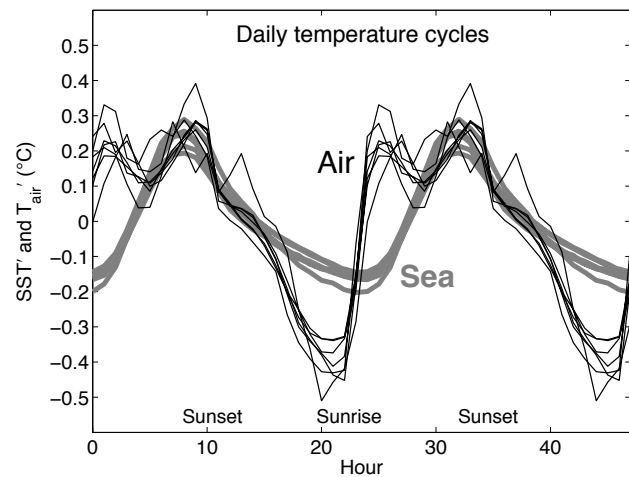

**Figure 2.** Observed daily cycles of sea surface temperature (SST') and surface air temperature ($T_{air}$'). Prime (') indicates deviation from the mean daily temperature (°C). Each line represents the daily cycle averaged over the autumnal solstice period (September–October) of seven different years: 1992–1998, inclusive. Raw hourly data from the NOAA Tropical Atmosphere Ocean (TAO) mooring at 155°W, 0°N (McPhaden *et al.* 1998). The amplitude of the daily cycle of surface air temperature here is roughly half of the average over Earth's tropical ocean regions (Dai and Trenberth 2004). These values are consistent with Kennedy *et al.* (2007).

Owing to theoretical developments in the 20th Century (Bjerknes 1969, Lindzen and Nigam 1987) along with modern observations and empirical models (Sun 2003), we know that temperature gradients translate into pressure gradients, and that, of course, pressure gradients ultimately set the wind in motion. From these principles, we can calculate roughly the wind speeds and directions that would result from the temperature variations born out in real measurements in Fig. 2 and shown schematically in Fig. 3.

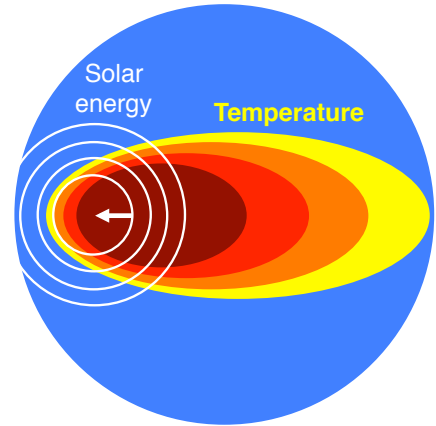

**Figure 3.** Map of incident solar energy and temperature on a simplified, ocean–dominated, non–rotating Earth in a solar system where the Sun revolves around the Earth daily (*i.e.*, a Ptolemaic universe). The symmetric patch of maximum solar energy on the daytime side of Earth (concentric circles) propagates westward. The *asymmetric* patch of warm surface temperature (colored ellipses) follows closely behind. Winds are proportional to horizontal gradients of pressure, which are coupled to horizontal gradients of temperature.

Given the strong east–west (and temporal) asymmetry in the temperature pattern, one would expect a narrow corridor of strong westerly winds on the leading edge of the sunlit side, followed by a broad swath of weaker but persistent easterly winds on the trailing side. This is indeed what is seen in a snapshot of the calculated winds (Fig. 4A), with the additional wind reversal at about the 150th meridian. The latter feature is related to the noticeable drop in air temperature in the otherwise warmest part due to the development of thunderstorms. In total, about 70% of the tropics at a given time would be experiencing the easterly trade winds, and only 25% experiencing westerlies (Fig. 4B).

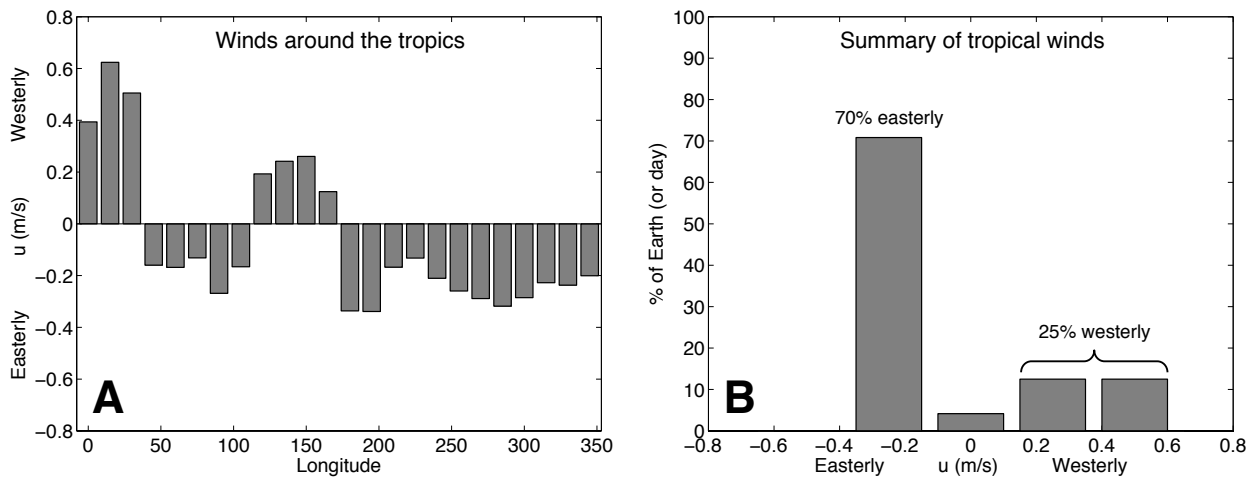

**Figure 4.** (**A**) Profile of the east–west component of the wind along a line of latitude in the tropics calculated from observed temperatures (*i.e.*, Fig. 2). Positive indicates westerly (from the west, or eastward) and negative indicates easterly. 1 m/s equals roughly 2 knots, and the average strength of the easterly trade winds on Earth is about 4–6 m/s (Kalnay *et al.* 1996). (**B**) Histogram of calculated winds showing that 70% of the tropics would experience easterly winds at a given point in time, while 25% would experience westerly winds. The percentages also apply to the amount of a day at a single, fixed location.

While the argument and calculations presented here are crude, and the resultant wind speeds reach only about 10% of their observed values, the notion of the "wrongness" of Halley's proposed explanation can be reconsidered when cast within the milieu of the prevailing cosmological model he endeavored as an astronomer to overturn. It has been suggested by some authors that Halley's ideas on atmospheric circulation were the first to "go beyond the Ptolemaic view of climate" (Edwards). Ironically, Halley's explanation for the easterly trades was essentially correct, but in a Ptolemaic (geocentric) universe.

## Acknowledgements

The author thanks Lonnie Lippsett for helpful suggestions on improving the manuscript, and Caroline Ummenhofer for many discussions about the Hadley Circulation.